\documentclass{iaus}
\usepackage{graphicx}

\title[Jet-disk coupling across the stellar mass scale] 
{Investigating accretion disk - radio jet coupling across the stellar mass scale}

\author[James Miller-Jones et al.]   
{James C.A.~Miller-Jones$^1$,  Gregory R.~Sivakoff$^2$, Diego Altamirano$^3$, Elmar G.~K\"ording$^4$, Hans A.~Krimm$^5$, Dipankar Maitra$^6$, Ron A.~Remillard$^7$, David M.~Russell$^3$, Valeriu Tudose$^8$, Vivek Dhawan$^9$, Rob P.~Fender$^{10}$, Sebastian Heinz$^{11}$, Sera Markoff$^3$, Simone Migliari$^{12}$, Michael P.~Rupen$^9$ \and Craig L.~Sarazin$^2$}

\affiliation{$^1$ICRAR - Curtin University of Technology, GPO Box U1987, Perth, WA 6845, Australia \\ email: {\tt james.miller-jones@curtin.edu.au} \\[\affilskip]
$^2$Department of Astronomy, University of Virginia, P.O. Box 400325, Charlottesville, VA 22904, USA \\email: {\tt grs8g@virginia.edu, sarazin@virginia.edu} \\[\affilskip]
$^3$Astronomical Institute `Anton Pannekoek', University of Amsterdam, P.O. Box 94249, 1090 GE Amsterdam, the Netherlands \\email: {\tt d.altamirano@uva.nl, s.b.markoff@uva.nl, d.m.russell@uva.nl} \\[\affilskip]
$^4$Universit\'e Paris Diderot and Service d'Astrophysique, UMR AIM, CEA Saclay, F-91191 Gif-sur-Yvette, France \\email: {\tt elmar@koerding.eu}\\[\affilskip]
$^5$NASA/Goddard Space Flight Center, Greenbelt, MD 20771, USA; and\\ USRA, 10211 Wincopin Circle, Suite 500, Columbia, MD 21044, USA\\email: {\tt hans.krimm@nasa.gov}\\[\affilskip]
$^6$Department of Astronomy, University of Michigan, Ann Arbor, MI 48109, USA \\email: {\tt dmaitra@umich.edu}\\[\affilskip]
$^7$MIT Kavli Institute for Astrophysics and Space Research, Building 37, 70 Vassar Street, Cambridge, MA 02139, USA \\email: {\tt rr@space.mit.edu}\\[\affilskip]
$^8$Netherlands Institute for Radio Astronomy, Oude Hoogeveensedijk 4, 7991 PD Dwingeloo, the Netherlands \\email: {\tt tudose@astron.nl}\\[\affilskip]
$^9$NRAO Domenici Science Operations Center, 1003 Lopezville Road, Socorro, NM 87801, USA \\email: {\tt vdhawan@nrao.edu, mrupen@nrao.edu}\\[\affilskip]
$^{10}$School of Physics and Astronomy, University of Southampton, Southampton SO17 1BJ, UK \\email: {\tt r.fender@soton.ac.uk}\\[\affilskip]
$^{11}$Astronomy Department, University of Wisconsin-Madison, 475. N. Charter St., Madison, WI 53706, USA \\email: {\tt heinzs@astro.wisc.edu}\\[\affilskip]
$^{12}$European Space Astronomy Centre, Apartado/P.O. Box 78, Villanueva de la Canada, E-28691 Madrid, Spain \\email: {\tt simone.migliari@sciops.esa.int}\\[\affilskip]

}

\pubyear{2010}
\volume{275}  
\pagerange{001--001}
\jname{Jets at all Scales}
\editors{G.E. Romero, R. Sunyaev \& T.M. Belloni, eds.}
\begin{document}

\maketitle

\begin{abstract}
Relationships between the X-ray and radio behavior of black hole X-ray
binaries during outbursts have established a fundamental coupling
between the accretion disks and radio jets in these systems.  I begin
by reviewing the prevailing paradigm for this disk-jet coupling, also
highlighting what we know about similarities and differences with
neutron star and white dwarf binaries.  Until recently, this paradigm
had not been directly tested with dedicated high-angular resolution
radio imaging over entire outbursts.  Moreover, such high-resolution
monitoring campaigns had not previously targetted outbursts in which
the compact object was either a neutron star or a white dwarf.  To
address this issue, we have embarked on the Jet Acceleration and
Collimation Probe Of Transient X-Ray Binaries (JACPOT XRB) project,
which aims to use high angular resolution observations to compare
disk-jet coupling across the stellar mass scale, with the goal of
probing the importance of the depth of the gravitational potential
well, the stellar surface and the stellar magnetic field, on jet
formation.  Our team has recently concluded its first monitoring
series, including (E)VLA, VLBA, X-ray, optical, and near-infrared
observations of entire outbursts of the black hole candidate
H1743-322, the neutron star system Aquila X-1, and the white dwarf
system SS Cyg.  Here I present preliminary results from this work,
largely confirming the current paradigm, but highlighting some
intriguing new behavior, and suggesting a possible difference in the
jet formation process between neutron star and black hole systems.

\end{abstract}

\firstsection 
\section{Introduction}

Jets are found in accreting systems throughout the visible Universe.  For stellar-mass accretors, the evolution of such jets occurs on human timescales, and can be probed by resolved monitoring observations.  X-ray binaries are one such class of objects, in which two distinct types of jets are observed, with a clear connection between the X-ray state of the source and the observed radio emission.  From the flat radio spectra seen in the hard X-ray state, the presence of steady, compact, partially self-absorbed outflows is inferred, which have been directly resolved in two black hole (BH) systems (\cite[Dhawan et al.~2000]{Dha00}; \cite[Stirling et al.~2001]{Sti01}). Brighter, optically-thin, relativistically-moving jets are associated with high-luminosity, soft X-ray states during outbursts (\cite[Mirabel \& Rodr\'\i guez 1994]{Mir94}).

\subsection{Black holes}
Our current understanding of the duty cycles and disc-jet coupling in black hole X-ray binaries (BH XRBs) derives from a compilation of X-ray spectral and timing information, together with radio flux density monitoring and a limited set of high-resolution radio imaging.  The current paradigm, or `unified model' (\cite[Fender et al. 2004]{Fen04}) suggests that the jet morphology and power correlate well with position in an X-ray hardness-intensity diagram (HID).  Steady, self-absorbed jets are inferred to exist in the very low luminosity quiescent state and the higher-luminosity hard state.  As the X-ray intensity increases in the hard state, so too does the jet power, with the radio and X-ray luminosities following a non-linear correlation, $L_{\rm Radio}\propto L_{\rm X}^{0.7}$.  At about $L_{\rm X} \sim 0.1$--$0.3 L_{\rm Edd}$, the X-ray spectrum begins to soften, and the jet speed increases as the inner disc radius moves inwards.  Below a certain X-ray hardness (the `jet line'), the core jet switches off and internal shocks develop in the flow, which are observed as bright, relativistically-moving radio ejecta.  The source may remain at high luminosity for several weeks, making repeated transitions back and forth across the jet line, before the X-ray luminosity eventually decreases to $\sim0.02L_{\rm Edd}$ where the spectrum hardens (note the hysteresis effect compared to the higher-luminosity hard-to-soft transition), the core jet is re-established, and the source fades back into quiescence.

\subsection{Neutron stars}

Of the many different classes of neutron star (NS) X-ray binaries, only the low-magnetic field systems have shown evidence for radio emission.  These systems are divided by mass accretion rate into two main classes; the Z-sources and the atoll sources, each with distinct X-ray spectral and timing characteristics.  The Z-sources are consistently accreting at or close to the Eddington luminosity, whereas the atolls are accreting at a somewhat lower level.  The X-ray spectral and timing properties of atolls show many similarities to black hole systems, with distinct soft (so-called `banana') and hard (`extreme island') X-ray states, making them the best sources to compare with black hole X-ray binary outbursts.

To date, only a handful of atoll sources have been detected in the radio band during simultaneous radio/X-ray observations, showing them to be systematically fainter than the black hole sources at the same Eddington-scaled X-ray luminosity \cite[(Migliari \& Fender 2006)]{Mig06}.  However, it appears that a similar correlation between radio and X-ray luminosities holds in the hard-state atoll sources, but with a lower normalization and a steeper power-law index; $L_{\rm R}\propto L_{\rm X}^{1.4}$.  At high X-ray luminosities, close to the Eddington limit where the sources show Z-type behavior, bright transient ejecta are thought to exist, just as in black hole systems, although there appears to be only mild suppression of radio emission in the atoll sources when they reach a soft X-ray state.

\subsection{White dwarfs}

A third class of interacting binaries where mass is transferred from a donor to a degenerate compact object has a white dwarf as the accretor. One class of such systems, the dwarf novae (a type of Cataclysmic Variable; CV), also have accretion discs, which periodically develop disc instabilities, leading to a sudden increase in the accretion rate and causing short-lived outbursts.  During such outbursts, which recur at intervals of several weeks, these systems brighten by several magnitudes in the optical band.  Despite similarities to XRBs, with accretion onto a compact object and outbursts triggered by disc instabilities, no jets have thus far been directly resolved in CVs.

Generalizing the HID to a `disc-fraction luminosity diagram' (DFLD) by plotting the optical flux of the system against the fraction of emission arising from the power-law spectral component (as opposed to disc emission) shows that outbursts of dwarf novae follow a very similar track to BH and NS systems.  Extending this analogy with the Unified Model then suggests that they should show flat-spectrum radio emission in the rise phase of an outburst, and resolved ejecta during the subsequent spectral softening.  The former prediction was spectacularly confirmed during an outburst of the dwarf nova SS Cyg (\cite[K\"ording et al. 2008]{Koe08}).  The radio emission was highly variable, peaking at 1.1\,mJy, and coincident with the optical outburst.  During the decay, the radio spectrum was slightly inverted, suggestive of a compact jet, as resolved in BH XRBs.  However, the existence of such a jet can only be directly verified with high-resolution imaging.

\section{The Jet Acceleration and Collimation Probe of Transient X-ray Binaries (JACPOT XRB)}

\subsection{Aims and strategy}

The Jet Acceleration and Collimation Probe of Transient X-ray Binaries (JACPOT XRB) project\footnote{http://www.astro.virginia.edu/xrb\_jets/} aims to probe the similarities and differences in the jet launching process between these three different source classes by conducting intensive monitoring of an outburst of each of these different classes of accreting compact object.  Time-resolved, high angular resolution monitoring in the radio band using the VLBA and (E)VLA, plus WSRT and EVN when available, allows us to directly image the evolution of the jets over the course of an outburst.  Simultaneous multi-wavelength coverage using {\it Swift} BAT, {\it RXTE} PCA, and {\it MAXI} in the X-ray band, plus optical and infrared monitoring using the two Faulkes Telescopes, FanCam and CTIO (via the SMARTS consortium), then enables us to tie the jet behavior to the corresponding changes in the accretion flow.

With such detailed, multi-wavelength monitoring, we aim to directly test the prevailing paradigm for black hole X-ray binary outbursts \cite[(Fender et al. 2004)]{Fen04} and, by comparing the jet behavior across the three different source classes, to ascertain the role played in jet formation by the depth of the gravitational potential well and the presence or absence of a stellar surface and stellar magnetic field.

\subsection{Status}

One outburst of each class of accreting compact source has now been observed as part of the JACPOT XRB project, and data reduction and analysis are underway.  Monitoring campaigns on one further black hole and one neutron star system have been approved by NRAO, and are awaiting triggering events.

\subsection{H1743-322}

\begin{figure}[t]
\begin{center}
 \includegraphics[width=\textwidth]{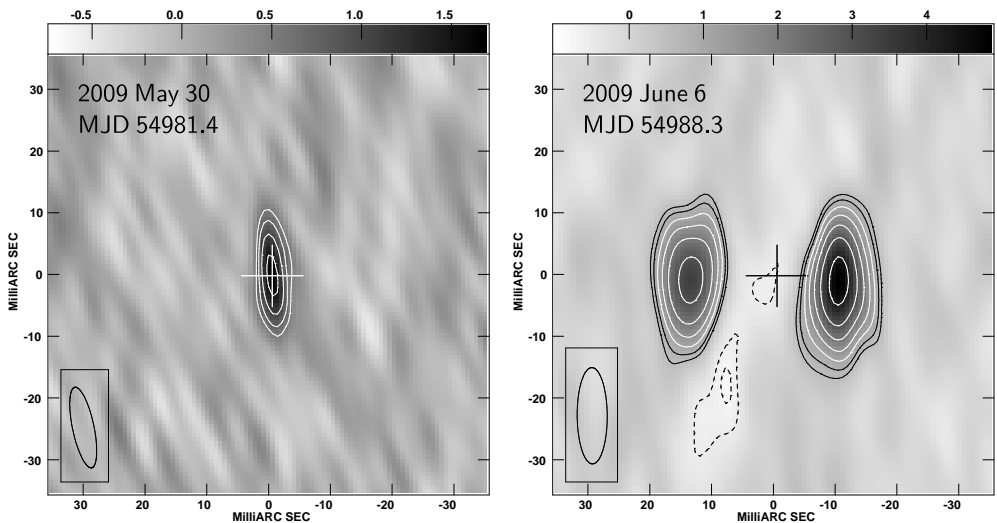} 
 \caption{VLBA images of H1743-322 at 8.4 GHz. An unresolved source consistent with a compact, steady jet is seen in the early epochs, with the core quenching and giving rise to bright, optically-thin ejecta after the transition to the soft intermediate state.  The grey scales denote flux density in mJy\,beam$^{-1}$, and the crosses mark the position of the central binary.}
   \label{fig:h1743_images}
\end{center}
\end{figure}

We triggered our first black hole observing campaign on the 2009 outburst of the black hole candidate source H1743-322.  The X-ray evolution of the outburst has already been analysed in detail by \cite[Motta et al. (2010)]{Mot10} and \cite[Chen et al. (2010)]{Che10}.  We monitored the outburst with the VLA, ATCA and VLBA, tracking the evolution of the radio emission through the entire outburst.  While the VLBA observations were hampered by the strong scattering and lack of good calibrators in the direction of the Galactic Center, we detected both the compact core during the initial rising hard state, and also the launching of ejecta immediately following the transition to the soft intermediate state (Fig.~\ref{fig:h1743_images}).

\begin{figure}[t]
\begin{center}
 \includegraphics[width=\textwidth]{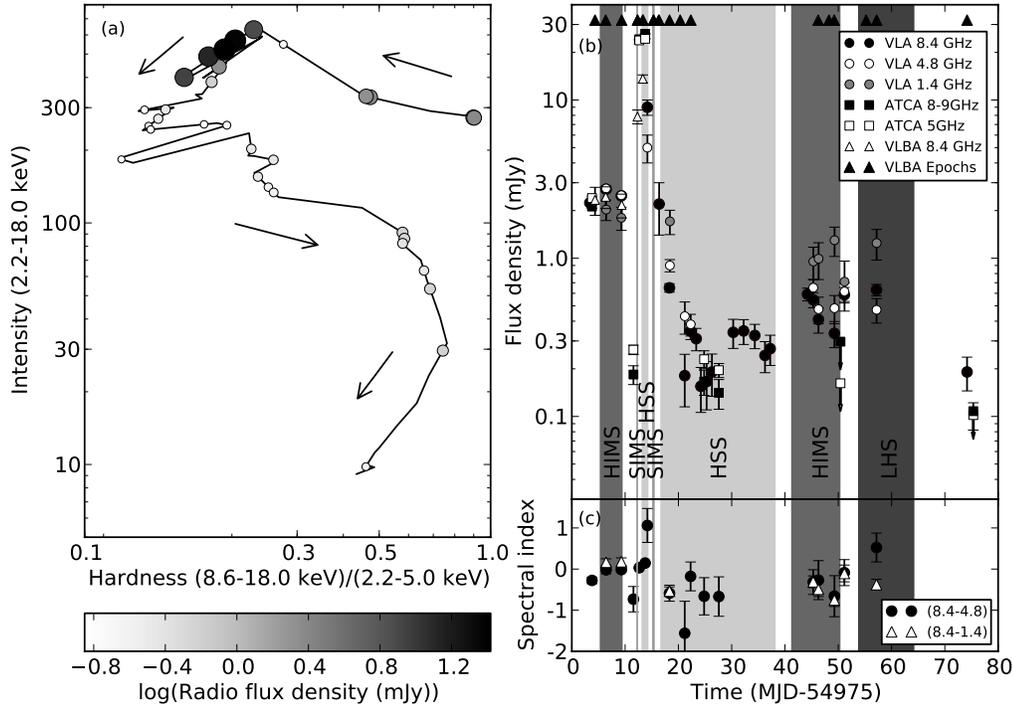} 
 \caption{Disc-jet coupling during the 2009 outburst of H1743-322. (a) HID for the outburst, with circles indicating radio detections, whose size and color indicate the measured radio flux density.  The X-ray state at the time of the radio observations has been interpolated from X-ray observations within 2\,d. Arrows show how the source evolves with time. (b) Radio light curve, showing the state classifications of \cite[Motta et al. (2010)]{Mot10}: HIMS = Hard Intermediate State; SIMS = Soft Intermediate State; HSS = High Soft State; LHS = Low Hard State.  (c) Radio spectral index as a function of time.}
   \label{fig:h1743_hids}
\end{center}
\end{figure}

The disc-jet coupling in this system follows the standard pattern outlined by \cite[Fender et al. (2004)]{Fen04}, as shown in the hardness-intensity diagram (HID) in Fig.~\ref{fig:h1743_hids}.  We see flat-spectrum radio emission from a compact, unresolved core jet in the hard intermediate state (HIMS) at the beginning of the outburst.  This core radio emission then quenches, giving rise to very faint, steep-spectrum emission as the source moves into the soft intermediate state (SIMS), following which we detect bright, optically-thin ejecta.  The radio emission does not completely vanish during the high soft state (HSS), but shows a modest increase in flux density once the source returns to the HIMS, before fading with time during the decay phase in the low hard state (LHS).  We draw particular attention to the pre-flare `quench' phase seen during the hard to soft transition.  Originally noted by \cite[Fender et al. (2004)]{Fen04}, this short-lived phase is often missed owing to limited temporal sampling in the radio band during outbursts of black hole X-ray binaries, but is surely key to understanding the nature of the transition from steady compact jets to bright, relativistically-moving ejecta.  Over the reverse transition, the reactivation of the compact jet occurs at a hardness ratio of $0.26<H<0.58$, and appears to support the assertion of \cite[Fender et al. (2009)]{Fen09} that the `jet-line' is not vertical.

\subsection{Aql X-1}

The 2009 November outburst of Aql X-1 was observed using the VLA, VLBA, e-EVN and {\it RXTE} PCA, resulting in the first milliarcsecond-scale VLBI detection of the source.  We obtained full spectral coverage of the outburst from the radio through infrared (FanCam, SMARTS), optical (Faulkes), ultraviolet ({\it Swift} UVOT) and X-ray bands.  Modeling of the full spectral energy distribution and its evolution is underway.

\begin{figure}[t]
\begin{center}
 \includegraphics[width=\textwidth]{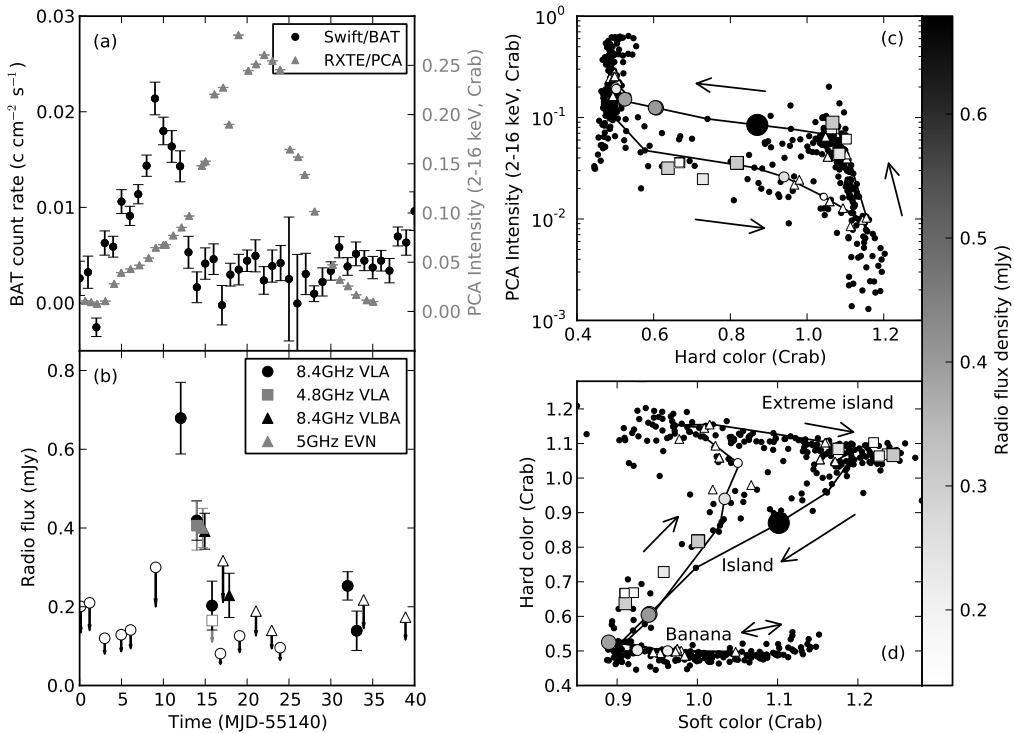} 
 \caption{The 2009 November outburst of Aql X-1.  (a) 15--50\,keV {\it Swift} BAT and 2--16\,keV {\it RXTE} PCA lightcurves.  (b) Radio lightcurves from the VLA, VLBA and EVN. Open markers denote upper limits at the corresponding array and frequency.  (c) HID showing the track of the 2009 November outburst, with dots representing all measured {\it RXTE} PCA points from previous outbursts.  Filled circles represent the radio measurements during the 2009 outburst and filled squares denote archival radio measurements, with the size and color of these markers corresponding to the measured radio flux density.  Hollow triangles denote non-detections.  Arrows show how the source evolves with time. (d) CCD of the 2009 outburst, with the same symbols as in (c).  Hard and soft colors are the count rate ratios (9.7--16.0\,keV/6.0--9.7\,keV) and (3.5--6.0\,keV/2.0--3.5\,keV) respectively, normalized by the Crab Nebula on a per-PCU basis.}
\label{fig:aqlx1_lcs}
\end{center}
\end{figure}

The radio and X-ray lightcurves, plus the hardness-intensity diagram (HID) and color-color diagram (CCD) of the outburst are shown in Fig.~\ref{fig:aqlx1_lcs}.  While full details of the observations and analysis may be found in \cite[Miller-Jones et al. (2010)]{Mil10}, we summarize the most salient points here.  While hysteresis in the HID of Aql X-1 is well known \cite[(Maitra \& Bailyn 2004)]{Mai04}, there have been few previously-reported radio detections \cite[(Tudose et al. 2009)]{Tud09}.  Our monitoring campaign provided the most complete radio coverage to date of an entire outburst of Aql X-1, finding the radio emission to be consistent with being activated by transitions from a hard spectral state to a soft state, and also by the reverse transition, just as seen in BH systems.  A further similarity with black hole systems was the quenching of the radio emission above a certain X-ray luminosity ($\sim10$\% of the Eddington luminosity) while in a soft spectral state.  However, in contrast to the BH systems, the VLBI observations combined with radio spectral information showed no evidence for steep-spectrum, optically-thin ejecta after the hard-to-soft state transition.  The radio emission was at all times consistent with a compact, partially self-absorbed jet as seen in the hard states of BH systems.  This may suggest a fundamental difference in the jet formation mechanism between the two classes of source.

\subsection{SS Cyg}

\begin{figure}[t]
\begin{center}
 \includegraphics[width=\textwidth]{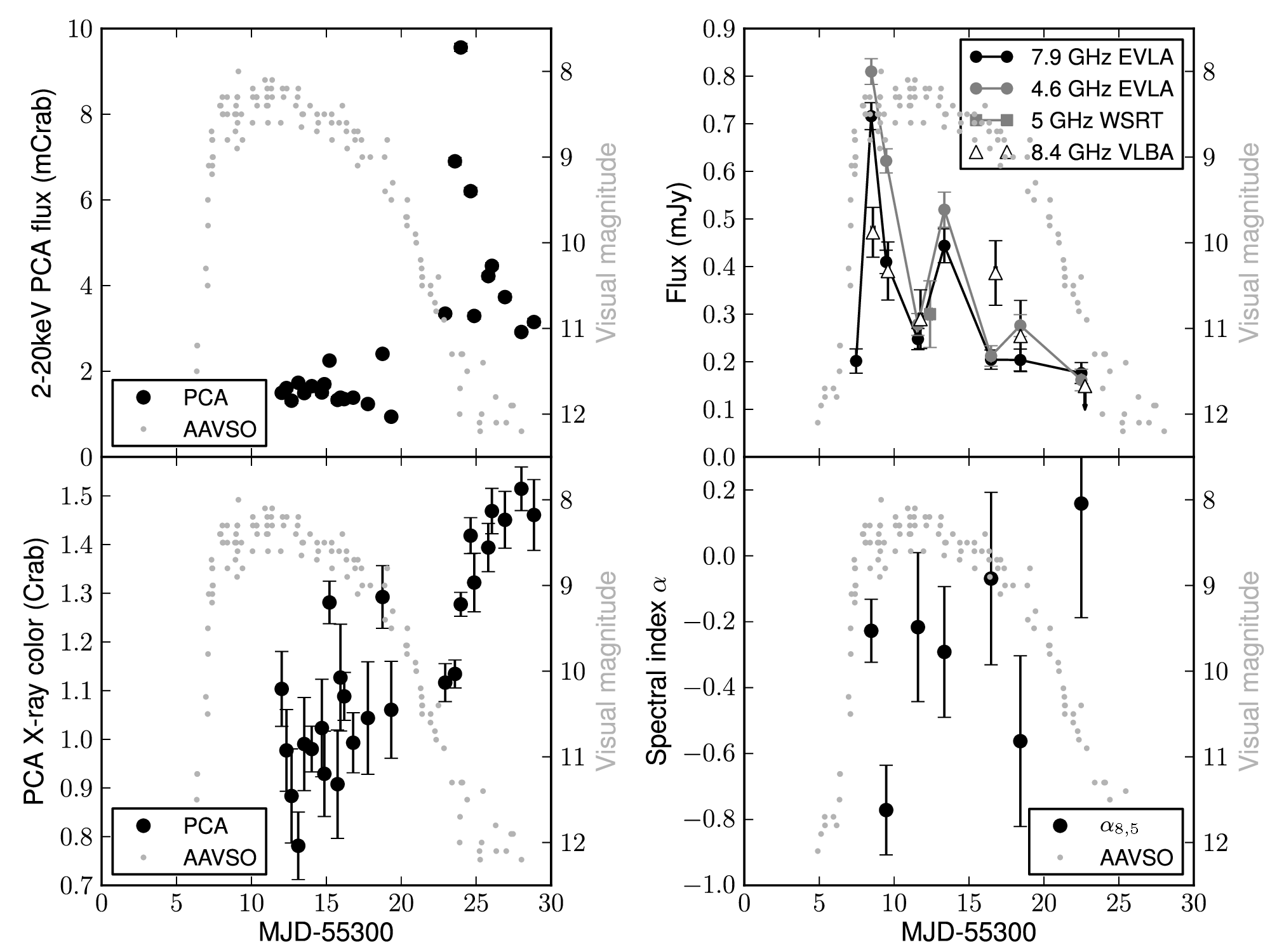} 
 \caption{Light curves of SS Cyg during the 2010 April outburst.  PCA
   X-ray color is the count rate ratio (6--16\,keV)/(2--6\,keV),
   normalized by the Crab nebula on a per-PCU basis.  Visual
   magnitudes from AAVSO are shown in light grey on each panel to mark
   the progress of the outburst.  The outburst evolution follows the
   standard pattern in the optical and X-ray bands.  EVLA radio
   detections were obtained at all epochs, and VLBA detections at the
   first 5 epochs.}
   \label{fig:sscyg_lcs}
\end{center}
\end{figure}

SS Cyg is the brightest dwarf nova, at a distance of only 166\,pc \cite[(Harrison et al. 1999)]{Har99}.  It undergoes several outbursts each year, and in 2010 April, the AAVSO notified us that a new outburst of the source was beginning.  We were able to get on source with EVLA within 24\,h, and followed up the ensuing detection with a full monitoring campaign using EVLA, WSRT, VLBA, {\it RXTE}, {\it Swift} and AAVSO.  Unfortunately the source is too bright for {\it Swift} UVOT to provide coverage in the ultraviolet.  The X-ray and radio light curves of the outburst are shown in Fig.~\ref{fig:sscyg_lcs}.  The outburst proceeded in a very similar fashion to that monitored by \cite[Wheatley et al. (2003)]{Whe03}, showing a sharp drop in the hard X-ray emission at the peak of the optical outburst, and a subsequent recovery as the optical emission faded.  As seen by \cite[K\"ording et al. (2008)]{Koe08}, the radio emission rose rapidly at the beginning of the outburst, showing a steep radio spectrum ($\alpha=-0.77\pm0.14$, with $S_{\nu}\propto\nu^{\alpha}$) immediately after the peak, suggestive of optically-thin ejecta.  During the decay phase later in the outburst, the observed flatter radio spectrum is more consistent with a partially self-absorbed compact jet.

The source was detected by the VLBA during 5 of the 6 epochs.  These detections imply a brightness temperature in excess of $5\times10^6$\,K.  As argued by \cite[K\"ording et al. (2008)]{Koe08}, together with the radio spectral constraints this suggests the presence of an unresolved jet, as seen in accreting neutron star and black hole systems.

As an aside, the astrometric accuracy of the VLBA observations allowed us to fit for the source proper motion over the 10 days of observation.  Assuming a distance of 166\,pc \cite[(Harrison et al. 1999)]{Har99}, we find a proper motion of $\mu_{\alpha}\cos\delta = 110\pm4$\,mas\,y$^{-1}$ in R.A., and $\mu_{\delta}=48\pm6$\,mas\,y$^{-1}$ in Dec., fairly consistent with that given in UCAC3 \cite[(Zacharias et al. 2010)]{Zac10}, but significantly different from the proper motion derived by \cite[Harrison et al. (2000)]{Har00}, with implications for the accuracy of the HST astrometric measurements.  A future revision of the distance could remove the discrepancy between predictions of the disc instability model and the observed outburst luminosity \cite[(Schreiber \& Lasota 2007)]{Sch07}.

\section{Summary}

We have embarked on an ambitious project to test the disc-jet coupling in three different classes of accreting stellar-mass compact objects, using high-resolution VLBI radio observations coupled with multiwavelength monitoring data.  Using these data, we aim to (a) test the `unified model' of \cite[Fender et al. (2004)]{Fen04} for the disc-jet coupling in black hole X-ray binaries, and (b) to determine the role played in jet formation by the depth of the gravitational potential well, the stellar surface and the stellar magnetic field.  We have already observed one outburst from each of a black hole, a neutron star and a white dwarf binary system, and while preliminary analysis of the data largely confirms the existing paradigm, some interesting similarities and differences between the evolution of black hole and neutron star outbursts have been found.

\section{Acknowledgements}
We are very grateful to the NRAO, {\it RXTE} and {\it Swift} schedulers for their flexibility and prompt responses which have made these observing campaigns feasible.  We would also like to thank Mickael Coriat and Stephane Corbel for sharing their ATCA data on H1743--322, and Matthew Templeton and the worldwide network of AAVSO observers who triggered and supported our observing campaign on SS Cyg.

\end{document}